\documentclass[aps,preprint,a4paper,showpacs,showkeys,superscriptaddress]{revtex4-1}
\ifx\pdfoutput\undefined
\usepackage[usenames]{color} 
\usepackage[dvips]{graphicx}
\else
\usepackage{color} 
\definecolor{BrickRed}{rgb}{0.85,0.15,0.25}
\definecolor{MidnightBlue}{rgb}{0,0.45,0.85}
\definecolor{ForestGreen}{rgb}{0,0.85,0.45}
\usepackage{graphicx}
\usepackage{epstopdf}
\usepackage{subfigure}
\usepackage{latexsym,amsmath,amssymb}
\allowdisplaybreaks

\usepackage{umoline}
\newsavebox\CBox

\begin{document}


\title{Thermodynamic phase transition in the rainbow Schwarzschild black hole}

\author{Yongwan Gim}%
\email[]{yongwan89@sogang.ac.kr}%
\affiliation{Department of Physics, Sogang University, Seoul 121-742,
  Republic of Korea}%

\author{Wontae Kim}%
\email[]{wtkim@sogang.ac.kr}%
\affiliation{Department of Physics, Sogang University, Seoul 121-742,
  Republic of Korea}%

\date{\today}

\begin{abstract}
We study the thermodynamic phase transition in the rainbow Schwarzschild black hole
where the metric depends on the energy of the test particle. Identifying
the black hole temperature with the energy from the modified dispersion
relation, we obtain the modified entropy and thermodynamic energy along with the modified local temperature in the cavity 
to provide well defined black hole states.
It is found that apart from the conventional critical temperature related 
to Hawking-Page phase transition there appears an additional critical temperature 
which is of relevance to the existence of a locally stable tiny black hole; however, the off-shell free
energy tells us that this black hole should eventually tunnel into the stable large black hole. 
Finally, we discuss the reason why the temperature near the horizon 
is finite in the rainbow black hole by employing the running gravitational coupling constant, 
whereas it is divergent near the horizon in the ordinary Schwarzschild 
black hole.

\end{abstract}



\maketitle

\section{Introduction}
\label{sec:intro}
There has been much attention to modified dispersion relations (MDR) which are associated with the energy of test particle in gravity's rainbow. Since they have given not only experimental explanation for the threshold anomalies 
in ultra high cosmic rays 
and Tev photons \cite{AmelinoCamelia:1997gz, AmelinoCamelia:1997jx, Colladay:1998fq, Coleman:1998ti, AmelinoCamelia:1999wk,AmelinoCamelia:2000zs, Jacobson:2001tu, Jacobson:2003bn}
but also have appeared  theoretically
in the semi-classical limit of loop quantum gravity \cite{Gambini:1998it, Alfaro:2001rb, Sahlmann:2002qk, Smolin:2002sz, Smolin:2005cz}, the rainbow gravity has been extensively studied 
in order to explore various aspects of black holes and cosmology 
\cite{Galan:2004st, Galan:2005ju, Hackett:2005mb, Aloisio:2005qt, AmelinoCamelia:2005ik, Ling:2005bp, Galan:2006by, Ling:2006az, Ling:2006ba, Amelino-Camelia:2013wha, Barrow:2013gia, Ling:2008sy,  Garattini:2011hy, Girelli:2006fw, Garattini:2011fs, Garattini:2011kp, Liu:2007fk, Peng:2007nj, Li:2008gs, Ali:2014xqa, Awad:2013nxa}.

While the ordinary uncertainty principle has been promoted to
 the generalized uncertainty principle (GUP) in the quantum regime \cite{Maggiore:1993kv, Kempf:1994su, KalyanaRama:2001xd, Chang:2001bm, Setare:2004sr, Medved:2004yu, Nozari:2005ex, Setare:2005sj, Ling:2005bq, Adler:2001vs},
 the MDR has been essentially based on the deformation of relativity called the doubly special relativity 
\cite{AmelinoCamelia:2000ge, AmelinoCamelia:2000mn, AmelinoCamelia:2003uc, AmelinoCamelia:2003ex, Magueijo:2001cr, Magueijo:2002am} 
which is the extended version of Einstein's special relativity 
in that the Plank length is also required to be invariant under any inertial frames
 apart from
the invariant speed of light.
However, it has been claimed that  in this framework a nonlinear Lorentz transformation in
the momentum space is needed 
to keep the double invariant constants accompanying a deformed Lorentz symmetry, so that
the ordinary dispersion relation should be modified by the nonlinear Lorentz transformation.
On the other hand, instead of this non-linear realization of Lorentz transformation,
Magueijo and Smolin  \cite{Magueijo:2002xx}
 proposed that the spacetime background felt by a test particle would depend on
its energy such that the energy of the test particle deforms the background geometry and
consequently the dispersion relation.

In particular, there have been extensive studies for black hole temperature
in rainbow black holes in connection with black hole thermodynamics \cite{AmelinoCamelia:2005ik, Ling:2005bp, Galan:2006by, Liu:2007fk, Ali:2014xqa, Li:2008gs}. 
In fact, the black hole temperature can be easily determined from the MDR and 
the uncertainty 
relation once the spacial uncertainty is identified with the size of the black hole
\cite{Adler:2001vs}.
Explicitly, from the uncertainty relation one can relate  
the particle momentum with the black hole mass  by identifying the position uncertainty with the 
black hole horizon, and then 
obtain the black hole temperature from the particle energy 
which is associated with the momentum through the dispersion relation.
On the other hand, it could also be defined straightforwardly by the surface gravity
 in such as the metric of the rainbow black hole, where it would naturally 
depend on the energy of the test particle \cite{Ling:2005bp}. 
Note that the latter temperature from the surface gravity should 
be consistent with the former one using the MDR and Heisenberg
uncertainty relation.
 Recently, the thermodynamic quantities were calculated in the rainbow
Schwarzschild black hole with a particular choice of rainbow functions 
and corresponding thermodynamics was studied in Ref. \cite{Ali:2014xqa}. However, 
the energy was directly regarded as the black hole temperature by invoking the ordinary dispersion relation rather than the MDR
in the rainbow gravity. 

In this work, we would like to reconsider this issue and calculate the black hole temperature 
from the definition of the surface gravity in
the rainbow Schwarzschild black hole and then the energy dependence of the temperature 
will be eliminated
by employing both the MDR and the uncertainty relation.
This procedure consistently determines the black hole
temperature and  gives rise to a certain different temperature from the
previous one in Ref. \cite{Ali:2014xqa}. 
So it would be interesting to study the thermodynamic quantities of entropy and heat capacity according to
this newly defined
 black hole temperature.
Furthermore, we shall investigate thermodynamic 
phase transition of the rainbow Schwarzschild black hole in order to find out how much the 
rainbow effect of the metric
 changes the ordinary phase transition in black hole thermodynamics. 

In section \ref{sec:temp}, the black hole temperature for the rainbow Schwarzschild black hole
will be calculated following the definition of the standard surface gravity, and then
the energy dependence of the test particle
in the rainbow metric will be rephrased by the use of two elements of the
MDR and the Heisenberg uncertainty principle.
 In section \ref{sec:thrm}, from the first law of thermodynamics,
the entropy is derived and then local thermodynamic quantities including
the thermodynamic energy and heat capacity in the cavity
 will be presented along the line of local thermodynamic approach in Ref. \cite{York:1986it}. 
In section \ref{sec:phase}, we shall obtain the on-shell free energy and the off-shell free energy 
for the rainbow black hole and
study phase transition between various black hole states and the hot
flat space. 
It will be shown that 
there exist two kinds of critical temperatures 
in contrast to the case of the ordinary Schwarzschild black hole.
A similar behavior appeared in the exactly soluble quantized Schwarzschild black hole \cite{Son:2012vj}.
Apart from the conventional critical temperature called the 
 Hawking-Page phase transition \cite{Hawking:1982dh, Cai:2007vv, Cai:2007wz,  Banerjee:2011au, Banerjee:2011raa}, there appears an additional critical temperature 
which has something to do with 
the existence of a locally stable tiny black hole; however, the off-shell free
energy shows that it should eventually tunnel into the stable large black hole. 
Finally, conclusion and discussion will be given  in section \ref{sec:discussion}.








\section{black hole temperature}
\label{sec:temp}
We are going to define the black hole temperature by employing the definition from
 the surface gravity of
the rainbow metric; however, it  depends on the energy of a test particle. So the main purpose in this
section is to show how to consistently 
eliminate the energy dependence from the temperature.    
Let us start with the MDR given in Ref. \cite{Magueijo:2002xx},
 \begin{equation}\label{MDR2}
 \omega^2 f(\omega/\omega_p)^2-p^2 g(\omega/\omega_p)^2=m^2,
 \end{equation}
where $\omega$,~$p$, $m$ are the energy,  momentum, mass  of a test particle, respectively, and 
the Planck energy is denoted by  $\omega_p$.
The functions of $f(\omega/\omega_p) , ~g(\omega/\omega_p)$ are so called the rainbow 
functions which will be determined depending 
on the specific models, where they should be reduced to $\lim_{\omega \rightarrow 0} f(\omega/\omega_p) = 1$ and~ $\lim_{\omega \rightarrow 0} g(\omega/\omega_p) = 1$ in the absence of the test particles. 

On the other hand, one of the interesting MDRs \cite{AmelinoCamelia:2008qg, AmelinoCamelia:1996pj}  
could be found  in the high-energy regime as
\begin{equation}\label{MDR1}
m^2 \approx \omega^2 - p^2 + \eta p^2\left(\frac{\omega}{\omega_p} \right)^n,
\end{equation}
where  $\eta$ is a positive free parameter and $n$ is assumed to be
positive integer.
Comparing  Eq. (\ref{MDR2}) with Eq. (\ref{MDR1}), one can determine the specific rainbow functions as \cite{Ali:2014xqa},
\begin{equation}\label{rainbowfunc}
f(\omega/\omega_p)=1, \quad  g(\omega/\omega_p)=\sqrt{1-\eta \left(  \frac{\omega}{\omega_p} \right)^n}.
\end{equation}
Next, let us consider the rainbow Schwarzschild black hole described as \cite{Magueijo:2002xx}
\begin{equation}\label{metric}
ds^2=-\frac{1}{f(\omega/\omega_p)^2}\left(1-\frac{2GM}{r}\right)dt^2+\frac{1}{g(\omega/\omega_p)^2}\left(1-\frac{2GM}{r}\right)^{-1}dr^2+\frac{r^2}{g(\omega/\omega_p)^2}
d\Omega^2,
\end{equation} 
then the Hawking temperature $T_{\rm H}$ is calculated as
\begin{equation}\label{HawkingT}
T_{\rm H}= \frac{\kappa_{\rm H}}{2\pi} = \frac{1}{8\pi G M}\frac{g(\omega/\omega_p)}{f(\omega/\omega_p)},
\end{equation}
where $\kappa_H$ is the surface gravity at the horizon. By the use of the 
explicit form of rainbow functions \eqref{rainbowfunc}, the black hole temperature can be written as 
\begin{equation}
\label{tem}
T_{\rm H}= \frac{1}{8\pi G M}\sqrt{1-\eta\left(\frac{\omega}{\omega_p}\right)^n}.
\end{equation}
In order to eliminate the dependence of the particle energy in the black temperature of the rainbow gravity
\eqref{tem}, 
one can use
 the Heisenberg uncertainty principle 
of $\Delta x \Delta p \sim 1$, which yields a relation between 
the momentum $p$ and the mass of  black hole $M$ \cite{Adler:2001vs} 
 \begin{equation}\label{deltap}
p= \Delta p \sim  \frac{1}{2 G M}
 \end{equation}
where $\Delta x =2 G M$.
In principle, plugging Eq. \eqref{deltap} into  the MDR \eqref{MDR1},
one can determine the energy of $\omega$; however, 
it is  non-trivial to solve the MDR for a general $n$. 
If we choose $n=2$ for simplicity, 
the energy for the massless particle can be easily 
solved as  
 \begin{equation}\label{omega2}
\omega =  \frac{\omega_p}{\sqrt{\eta+4 G^2 \omega_p^2 M^2}}.
\end{equation}
Plugging the above relation \eqref{omega2} into the temperature (6) for $n=2$, we can obtain 
new temperature as
\begin{equation}\label{T2}
T_{\rm H}=  \frac{1}{8\pi G M}\sqrt{\frac{4G M^2}{4G M^2+\eta}},
\end{equation}
with $G=1/\omega_p^2$. 
It goes to asymptotically well-known Hawking temperature
while it is finite for $M \to 0$.
Thus, the temperature \eqref{tem} of the rainbow black hole could be expressed  
in terms of the mass of the black hole by solving both the 
MDR \eqref{MDR1} and the Heisenberg uncertainty relation \eqref{deltap}.

Now one might wonder what the difference is between our result \eqref{T2} and 
the previous result in Ref. \cite{Ali:2014xqa}, where
the temperature was given as 
 \begin{equation}\label{AliT}
T'_{\rm H}=\frac{1}{4\pi (2GM)^\frac{n+2}{2}}\sqrt{(2GM)^n-\frac{\eta}{\omega^n_p}}.
\end{equation}
In this case,  the mass of the black hole has  a lower bound of $M={\eta}^{1/n}/(2G \omega_p )$
due to the negative sign in the square root.
In Ref. \cite{Ali:2014xqa},
the particle energy was regarded as the particle momentum 
such as $\omega=p= \Delta p \sim  1/(2 G M)$ for a massless particle and then 
this relation was plugged into Eq. \eqref{tem} directly in order to eliminate the $\omega$-dependence.
In other words, in Ref. \cite{Ali:2014xqa}
the MDR was partially used in the sense that
the author used the correct  definition \eqref{tem} based on the MDR 
in the surface gravity but subsequently 
the ordinary dispersion relation rather than the MDR was used in the course of eliminating 
the $\omega$-dependence by taking $\omega=p$ along with the
uncertainty relation \eqref{deltap}.
Therefore, for the consistent treatment, 
 we used  the MDR \eqref{MDR1} along with the uncertainty relation \eqref{deltap}
when the energy dependence was eliminated in Eq. \eqref{tem}. 

Note that for the case of highly massive black holes or in the
absence of the rainbow effect implemented by $\eta=0$, the temperature (\ref{T2}) is reduced to the ordinary Hawking temperature.
The most interesting thing to distinguish from the previous results in 
Ref. \cite{Ali:2014xqa}
 is that the temperature \eqref{T2} subject to the MDR has a massless limit rather than
the remnant whose mass would be 
$M_{\rm rem}=\sqrt{\eta}/(2 \sqrt{G})$ for $n=2$ as seen from the temperature \eqref{AliT}.

Although the massless limit is allowed in our calculations, 
there still exists the lower bound of the energy as
$\omega=\omega_{p}/\sqrt{\eta}$ as seen from Eq. \eqref{omega2}.
Thus, for $M \to 0$, the temperature  (\ref{T2}) becomes finite while it is divergent
in the ordinary Schwarzschild black hole. It implies that
 the divergent ordinary temperature could be regularized in 
the regime of the rainbow gravity  so that the parameter $\eta$ 
plays a role of the cutoff and it becomes finite as $T_{\rm H} = 1/(4\pi \sqrt{G\eta})$
where it is still divergent if the cutoff  is removed as $\eta \to 0$. 
In fact, it has also been shown that the divergent quantities such as entropy and free energy  
can be regularized by the use of 
the MDR  in the brick wall method in Ref. \cite{Garattini:2009nq}.
Thus the regularized finite behavior of the temperature would be better than the divergent one, 
which will be used in studying thermodynamic quantities in the next section. And, 
we will discuss whether the vanishing temperature
can be realized in this rainbow Schwarzschild  black hole or not in the last section. 

\section{thermodynamic quantities}
\label{sec:thrm}
In this section, we calculate  thermodynamic quantities in the rainbow Schwarzschild black hole \eqref{metric}
characterized by the rainbow functions (\ref{rainbowfunc}) fixed as $n=2$ for exact solubility.
For this purpose, the first law of black hole thermodynamics is required to obtain the black hole entropy,
and then the local temperature for hot black holes in a cavity \cite{Tolman:1930zza} 
will be derived in order for
the local thermodynamic energy and heat capacity. The local thermodynamic analysis given by York 
\cite{York:1986it} presents a well-defined thermodynamic partition function
which has something to do with the existence of the large stable black hole.


\begin{figure}[pb]
  \begin{center}
\subfigure[{~ local temperature}]{
  \includegraphics[width=0.48\textwidth]{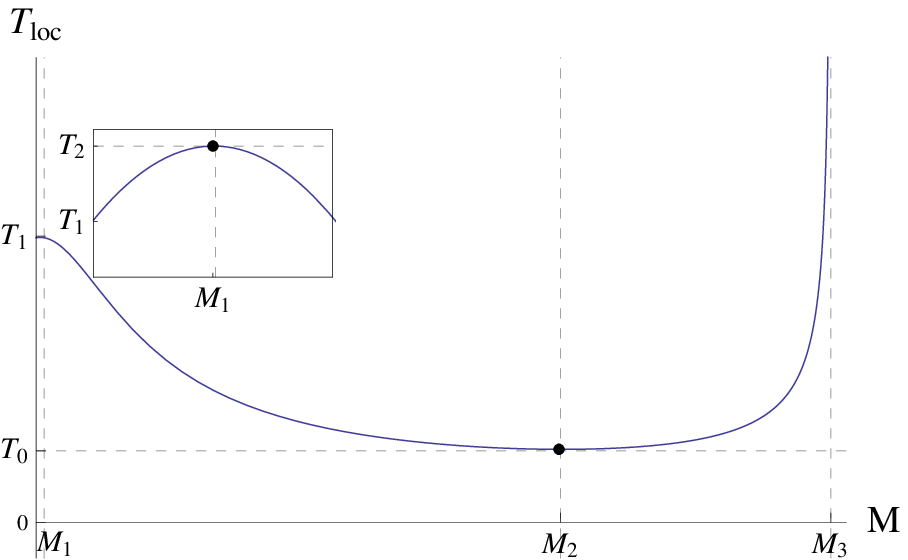}\label{fig:locT}}
\subfigure[{~ heat capacity}]{
  \includegraphics[width=0.48\textwidth]{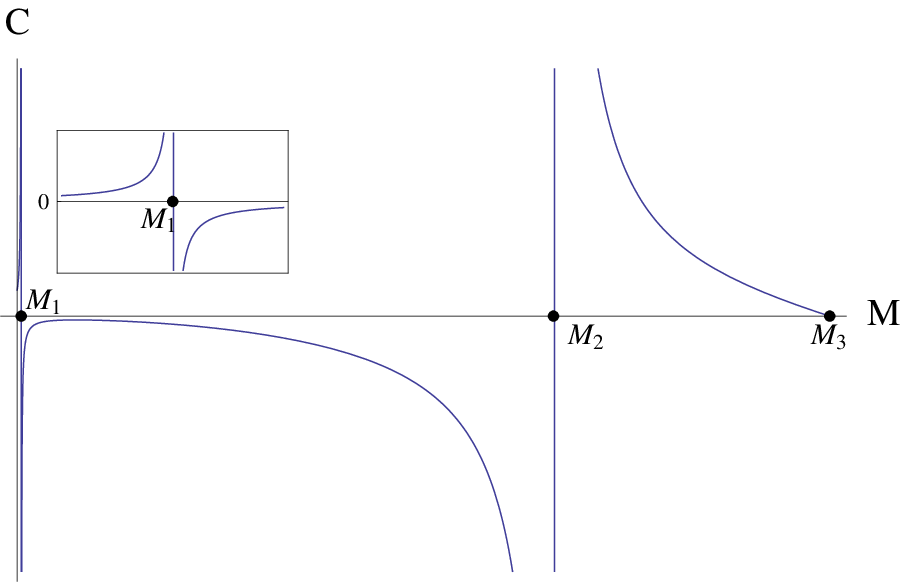}\label{fig:cap}}
  \end{center}
  \caption{The local temperature (a) and heat capacity (b) are plotted for $\eta=1,~r=10,$ and $G=1$. 
They show that the temperature have two extrema at $M_1$  and $M_2$, respectively in Fig. (a),
and the stability changes appear at those extrema in Fig (b). The maximum mass of the 
black hole is $M_3 =r/(2G)$.}
\end{figure}


From the first law of black hole thermodynamics of $dS=dM/T$, 
the entropy associated with the temperature (\ref{T2}) can be obtained as
\begin{equation}\label{S}
S=4\pi G M^2\sqrt{1+\frac{\eta}{4G M^2}}+\eta \pi \sinh^{-1}\left(\frac{2\sqrt{G}M}{\sqrt{\eta}}\right).
\end{equation}
For $\eta=0$, the entropy (\ref{S}) respects one-quarter of area law of $S=A/4$, where $A$ is 
the area of the black hole. 
The next leading order of correction to the area law is the logarithmic term as $S \approx A/4+\eta \pi/2 \ln(A/4)$,
which is reminiscent of quantum correction to the entropy \cite{Das:2001ic, Chatterjee:2003uv, Wang:2008zzc, Cai:2001dz, Myung:2010dv}.
In some sense, it is interesting to note that the rainbow metric plays a role of the quantum corrected metric. 
 
Now, the local temperature $T_{\rm{loc}}$ calculated at a finite distance $r$ outside the black hole 
is defined as \cite{Tolman:1930zza}, 
\begin{equation}\label{locT}
 T_{\rm{loc}} = \frac{1}{8\pi G M \sqrt{1-\frac{2G M}{r}}}\sqrt{\frac{4G M^2}{4G M^2+\eta}},
\end{equation}
where it is implemented by the redshift factor of the metric. 
For a given $r$,  the local temperature (\ref{locT})  is divergent as seen from Fig. \ref{fig:locT}
when the black hole size approaches $M_3$ where $r=2 G M_3$ since the black hole is very hot near the horizon, whereas 
it is finite as $T_1  = 1/(4\pi \sqrt{G\eta})$ for $M=0$ as was discussed 
in the previous section.
As for the black hole states, there are two extrema: one is the local minimum of $T_0$ at $M=M_2$ and the other
is the local maximum of $T_2$ at $M=M_1$. Note that there are two
black hole states for $T_0< T < T_1$ and three black hole states 
for $T_1 <T <  T_2$,
while there appears just one black hole state for $T>T_2$
. The details will be studied in the next section 
together with analysis of the free energy. 

Let us calculate the local thermodynamic energy $E_{\rm{tot}}$ 
by employing the local thermodynamic
first law, which yields 
\begin{align}\label{Etot}
E_{\rm{tot}} &=\int^{M}_{0} T_{\rm{loc}} dS  \notag \\
&=\frac{r}{G}\left(1-\sqrt{1-\frac{2G M}{r}}\right),
\end{align}
where we used the entropy \eqref{S} and the temperature \eqref{locT}. 
It happens to be the same with conventional expression which
is independent of the energy of test particles since 
the choice of rainbow functions \eqref{rainbowfunc} shows $f=1$ so that
the time-like Killing vector is the same with the ordinary one.
To investigate thermodynamic stability of the black hole, we
calculate the heat capacity defined as
 $C=\partial E_{\rm{tot}}/\partial T_{\rm{loc}}$, and explicitly it reads
\begin{equation}\label{capacity}
C=\frac{4\pi (r-2GM)(4GM^2+\eta)^\frac{3}{2}}{\sqrt{G}(12GM^2-4M r+\eta)}.
\end{equation}
For convenience's sake, let us define the black hole states 
depending on its mass scale as
tiny, small, and large lack hole satisfying $M< M_1$, $M_1<M <M_2$, and
$M_2<M <M_3$, respectively.
As seen from Fig.\ref{fig:cap}, there are two stable regions of
$M<M_1$ and $M_2<M<M_3$, while there is only one unstable region of $M_1<M<M_2$,
and thus the tiny and large black hole are stable and the small black hole is unstable.
Note that there was no stable tiny black hole in the ordinary Schwarzschild black hole \cite{York:1986it}.
What needs to be answered is that whether these locally stable states would undergo tunneling or not 
by investigating the free energy, which is studied in the next section.



\section{Free energy and phase transition}
\label{sec:phase}

\begin{figure}[pt]
  \begin{center}
\subfigure[{~Schwarzschild black hole}]{
  \includegraphics[width=0.48\textwidth]{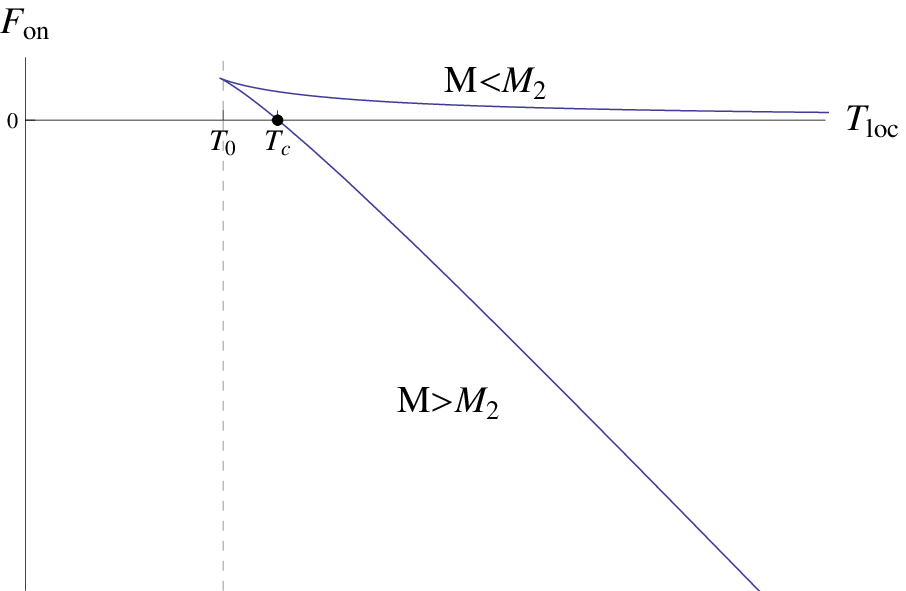}\label{fig:FonTsch}}
\subfigure[{~Rainbow black hole}]{
  \includegraphics[width=0.48\textwidth]{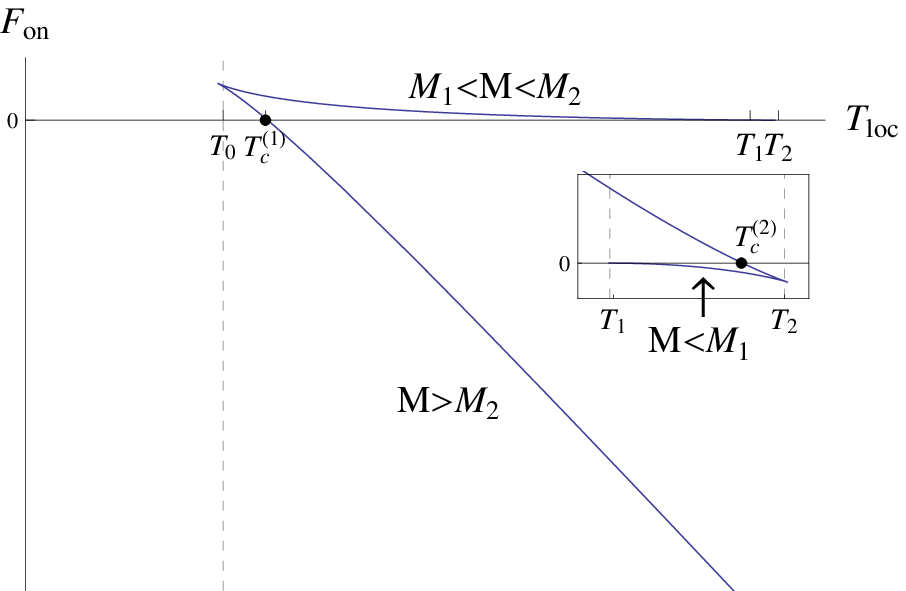}\label{fig:FonT}}
   \end{center}
  \caption{The on-shell free energy which is a function of the local temperature is plotted by setting $\eta=1, r=10,$ and $ G=1$. There exists a single critical temperature 
  $T_c$ in the   ordinary Schwarzschild black hole in Fig. 2 (a), while
  there appear two critical temperatures of $T_{c}^{(1)}$ and $T_{c}^{(2)}$ as in Fig. 2 (b).}
\end{figure}
In this section, we first calculate the on-shell free energy 
defined as $F_{\rm on}=E_{\rm tot}-T_{\rm loc}S$ 
in the rainbow Schwarzschild black hole in order to study thermodynamic phase transition \cite{York:1986it,Hawking:1982dh, Son:2012vj, Cai:2001dz, Myung:2010dv}. 
As is well known in the ordinary Schwarzschild black hole, the hot flat space is more probable than 
the large stable black hole below $T_c$, while the large stable black hole is 
more probable than the hot flat space above  $T_c$.
Moreover, the on-shell free energy of the unstable small back hole of $M<M_2$ 
is always positive as seen from Fig. \ref{fig:FonTsch}, so that the small black hole
should decay into the hot flat space as long as $T> T_0$.
\begin{figure}[pt]
  \begin{center}
\subfigure[{~ $T_0<T<T_1$}]{
 \includegraphics[width=0.48\textwidth]{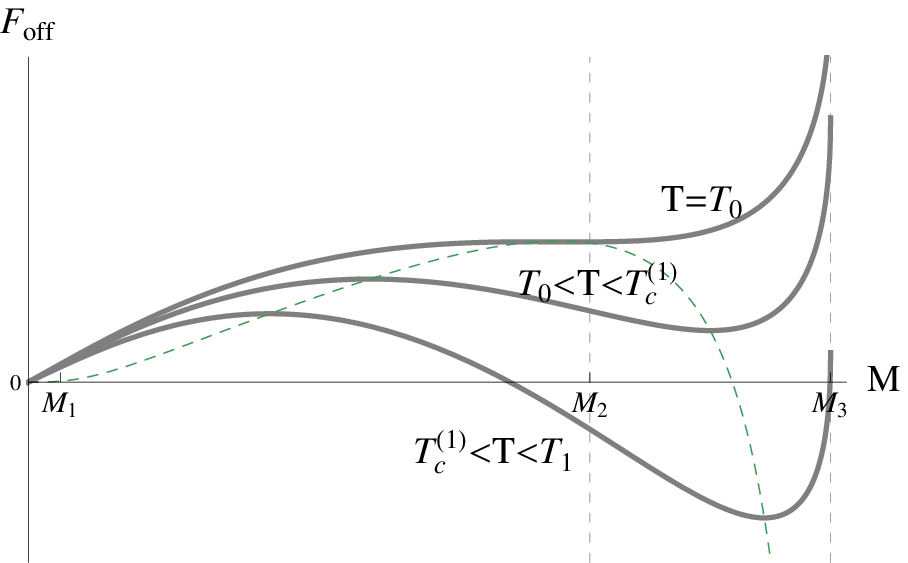}\label{fig:Foff1}}
\subfigure[{~ $T_1 <T<T_2$}]{
 \includegraphics[width=0.48\textwidth]{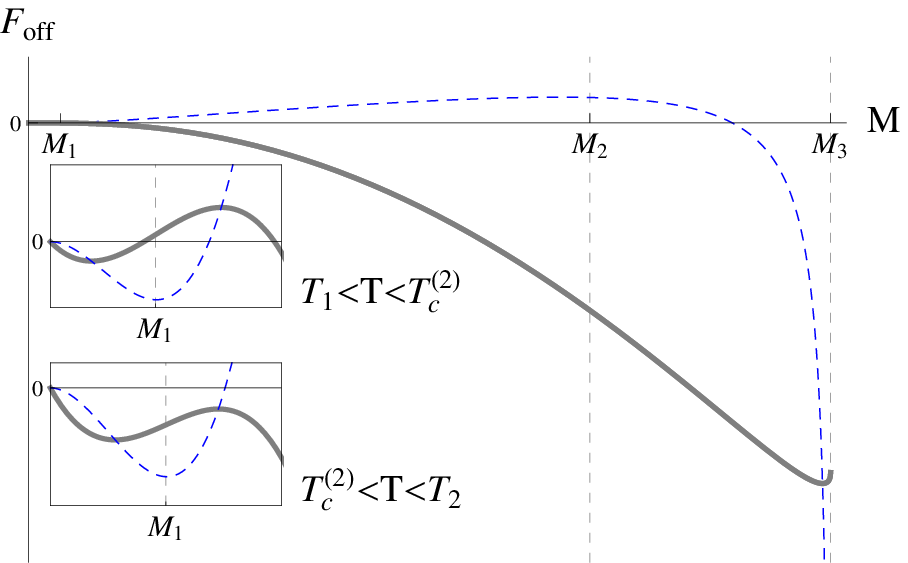}\label{fig:Foff4}}
   \end{center}
  \caption{The off-shell free energy subject to a temperature is plotted by setting $\eta=1, r=10,$ and $ G=1$, and it is
   expressed by a solid curve. The dotted curve is for the on-shell free energy which corresponds to the extrema of
 off-shell free energy. In Fig. (b),  the two small boxes are presented in order to explicitly show the off-shell free energy around the second critical temperature $T_c^{(2)}$.}
  \label{fig:F} 
\end{figure}
In the rainbow Schwarzschild black hole, the on-shell free energy  is also
calculated  by the use of
the temperature (\ref{locT}) and 
the thermodynamic energy \eqref{Etot}, which is explicitly written as 
\begin{equation}\label{Fon}
F_{\rm{on}}=\frac{r}{G}\left(1-\sqrt{1-\frac{2G M}{r}}\right)-\frac{1}{\sqrt{1-\frac{2G M}{r}}}\left(\frac{M}{2}+\frac{\eta \sinh^{-1}\left(\frac{2\sqrt{G}M}{\sqrt{\eta}}\right)}{4\sqrt{4G^2M^2+G\eta}}\right),
\end{equation}
where it recovers the on-shell free energy of the Schwarzschild black hole 
for $\eta =0$ as expected. And, the on-shell free energy of the hot flat space
vanishes since $E_{\rm tot}=S=0$ as seen from Eq. \eqref{S} and the
integral form of the energy \eqref{Etot} for
$M \to 0$ in any arbitrary temperature, i.e., $F_{\rm on}^{\rm hfs}=0$.
In fact, this is also consistent with the result from Eq. \eqref{Fon} when we take
$M \to 0$ since $sinh^{-1}x \sim x$ for $x \rightarrow 0$.

Now, let us discuss mainly three regions  in Fig. \ref{fig:FonT}.
(i) For $T_0 <T <T_1$, the behavior of the free energy is analogous to the
conventional one as shown in Fig. \ref{fig:FonTsch}, so that the first critical 
temperature $T_c^{(1)}$ in the rainbow black hole plays a role of $T_{(c)}$
in the ordinary black hole.
(ii) For $T_1<T<T_c^{(2)}$, the hot flat space collapses to 
form a tiny black hole of $M< M_1$ which is stable as was shown from the positive heat capacity
in Fig. \ref{fig:cap}. (iii)  For $  T_c^{(2)}<T<T_2$, the hot flat space would collapse to the
small black hole of $ M_1 <M<M_2$.
As shown in Fig. \ref{fig:FonT}, the temperature for the tiny and small black holes
should be terminated at $T_2$,
and only the stable large black hole exists above $T> T_2$.  
From (ii) and (iii), the on-shell free energy of the tiny black hole 
is still higher than that of the large black hole, so that the tiny black hole undergoes a tunneling and
eventually decays into the large black hole. This tunneling effect can be easily 
understood in terms of the following off-shell
free energy.

By using the entropy (\ref{S}) and the energy (\ref{Etot})
with an arbitrary temperature,
the off-shell free energy defined as $F_{\rm off}=E_{\rm tot}-T S$ 
can be plotted in Fig. \ref{fig:F} (a) and (b).
In Fig. \ref{fig:F} (a), the overall behaviors of the off-shell free energy are coincident with those of
the ordinary Schwarzschild black hole in Ref. \cite{York:1986it} and those in the 
anti-de Sitter Schwarzschild black hole in Ref. \cite{ Hawking:1982dh} as long as $T_0 <T<T_1$
in that the large black hole tunnels into the hot flat space.  
However, as the temperature is increased, there appears a tiny black hole 
 above $T_1$ and  it decays into the large black hole across the potential barrier 
with the tunneling probability given as
the difference between the free energy of the unstable small black hole and that of the tiny black hole.
Note that the difference between the small and tiny free energies are always positive, since the free energy of the small black hole is always higher than that of
the tiny black hole
even although the free energy of the 
small black hole is positive in the upper box while it is negative in the lower box
depending on the temperature  as shown in Fig. \ref{fig:F} (b).

\section{Discussion}
\label{sec:discussion}
We have calculated local thermodynamic quantities in the rainbow Schwarzschild 
black hole subject to the MDR and study its phase transition in terms of investigating the on-shell and
off-shell free energies. First of all,
the momentum of the emitted particle from the black hole 
was expressed by the black hole mass based on the Heisenberg
uncertainty principle and then the temperature
was derived by employing the nontrivial dispersion relation 
between the energy and the momentum characterized by the specific MDR.
According to this modified black hole temperature, we considered the on-shell free energy and the off-shell
free energy of the black hole in the finite size by introducing 
the isothermal surface of the cavity. 
In contrast to the conventional Schwarzschild black hole, it was shown that 
there exists an additional stable tiny black hole 
together with the conventional black hole states above $T_1$.  
Apart from the well-known critical temperature in Hawking-Page phase transition, there exists an additional critical temperature 
which is of relevance to the existence of a locally stable tiny black hole; however, the off-shell free
energy shows that it should eventually tunnel into the stable large black hole
with the finite transition probability since this tiny black hole is just locally stable. 

The temperature in the rainbow Schwarzschild black hole with the rainbow functions
 \eqref{rainbowfunc} becomes finite
when the black hole evaporates completely, whereas it was divergent 
in the ordinary Schwarzschild black hole. The reason for this finiteness of the temperature 
is related to the fact that  the Newton constant is running 
as $G(\omega)=G/g(\omega)$ \cite{Magueijo:2002xx},
which can be read off from the rainbow metric \eqref{metric}. 
In our case, it can be explicitly expressed as $G(M)=G \sqrt{ 1+{\eta}/({4 G M^2}) }$ by 
the use of Eq. \eqref{omega2}, 
then the temperature \eqref{T2} can be effectively 
written as $T_{\rm H}(M)={1}/(8\pi G(M) M)$. 
For $M \to 0$, the gravitational coupling becomes strong at the order of $1/M$ that 
the temperature becomes finite in this rainbow black hole.
Then, one might wonder whether the black hole temperature in the rainbow gravity 
can be zero or not  for $M \rightarrow 0$  when we choose 
different types of rainbow functions. 
To answer this question, let us redefine arbitrary two rainbow functions as  
$f(\omega/\omega_p)=1 +\tilde{f}(\omega/\omega_p),~g(\omega/\omega_p)=1 +\tilde{g}
(\omega/\omega_p)$ for convenience, 
where $\lim_{\omega \rightarrow 0} \tilde{f}(\omega/\omega_p) =0 $ and 
$\lim_{\omega \rightarrow 0} \tilde{g}(\omega/\omega_p) =0 $. 
Without loss of generality, one can solve MDR in Eq. \eqref{MDR2} for the massless case
 and Heisenberg relation \eqref{deltap}, then it can be shown that
 $\omega(1+\tilde{f}(\omega/\omega_p))/(1+\tilde{g}(\omega/\omega_p)) = 1/(2GM)$.
Requiring the condition of $T \propto \omega \rightarrow 0$ with $ M \rightarrow 0$, 
it gives rise to inconsistency in the sense that the right hand side is divergent but the left hand side
is zero. Thus, any choices of rainbow functions
in the rainbow Schwarzschild black hole can not make the 
temperature vanish when $M \to 0$.

In the above discussion, it was shown that 
the ordinary Heisenberg uncertainty relation and the
MDR in the rainbow gravity did not make the temperature vanish for $M \to 0$.
Now, in order to get the vanishing temperature at the vanishing limit
of the black hole mass,
 one might try to consider some other ways such as
 (i) GUP consideration  \cite{Maggiore:1993kv, Kempf:1994su, KalyanaRama:2001xd, Chang:2001bm, Setare:2004sr, Medved:2004yu, Nozari:2005ex, Setare:2005sj, Ling:2005bq, Adler:2001vs}
 or (ii) 
an iteration 
procedure on Eq.  \eqref{MDR1} which amounts 
to applying iteration procedure to the
temperature \eqref{tem} directly. 
(i) For the first case,  the most simple 
modification of the uncertainty principle is realized as 
$  \Delta x \Delta p \ge  1 +   \ell^2 (\Delta p)^2 $, 
where it leads to the minimal length of $\Delta x_{\rm min} =
2\ell$. The cutoff $\ell$ can be chosen as the Planck scale 
so that it can be fixed as $\ell^2 =1/(\omega_p)^2$ to make our notations consistent. 
By setting $\Delta x =2 G M$ \cite{Adler:2001vs},
the modified temperature 
improved by using both the MDR for $n=2$ 
and the GUP can be obtained as  
$T=\left(\omega_p/4\pi\right)\left[\left(2M^2-(1-\eta)\omega_p^2-2M\sqrt{(M-\omega_p)(M+\omega_p)}
\right)\left(4\eta M^2+(1-\eta)^2\omega_p^2\right)^{-1}\right]^{1/2}$
which is reduced to the well-known temperature based on the conventional GUP,
which is 
 $ T_{GUP} = \left(M/4\pi\right)\left [ 1 - \sqrt{1  - \omega^2_p/ M^2}~\right]$
\cite{Adler:2001vs} for $\eta =0$. Unfortunately, 
the modified temperature is still finite as  $T=\omega_p/(4\pi \sqrt{1+\eta})$ at 
the minimum mass of $M=\omega_p$. This fact is not surprising since the finite temperature with the remnant is somehow a generic feature in the regime of the GUP \cite{Adler:2001vs}. 
Thus, the above GUP combining the MDR does not likely
to give the desired result. (ii) Secondly, 
apart from the first case subject to the GUP and the MDR,
 let us solve directly the temperature \eqref{tem} by taking advantage of
the relation in Ref. \cite{Ling:2005bp}.  
In fact, the particle energy has been regarded as the
temperature of the particle, and it is natural to
take the relation of 
$\omega=4\pi T$ with the calibration factor \cite{Adler:2001vs}. Putting 
this energy-temperature relation into Eq. \eqref{tem} directly, we can derive the equation, 
$\eta(4\pi/\omega_p)^nT^n+(8\pi G M)^2 T-1=0$ \cite{Ling:2005bp}.
Interestingly for $n=2$, the solution to this equation is
exactly coincident with the temperature \eqref{T2}.
The temperatures corresponding to other values of 
$n=1, 3,4$ have a single real positive definite solution respectively. 
The behaviors for $n=1,3,4$ near $M \to 0$ are very similar to the case of $n=2$
which gives the finite result as was shown in section II.
Note that the above equation was solved for $n \le 4$ in order for exact
solutions without numerical methods. 
We did not find any evidence to make the temperature vanish  
when $M \to 0$ even in the second case either.
Therefore, it deserves further study in this direction. 
 



\acknowledgments 
We would like to thank M. Eune and E. J. Son for exciting discussions.
W. K. was supported by the Sogang University Research Grant of 2014 (201410019).


\end{document}